\documentclass{elsart}
\usepackage{epsf}
\begin{document}        

\begin{frontmatter}

\begin{flushright}
SLAC-PUB-8239\\
September 1999 \\
\end{flushright}
\title{
Current Performance of the SLD VXD3
}

\author{Toshinori Abe}
\address{
Stanford Linear Accelerator Center, Stanford University,
Stanford, CA  94309
}

\collab{Representing the SLD Collaboration}

\hskip 1in

\begin{center}
\small{
Presented at 8th International Workshop on Vertex Detectors (VERTEX 99),\\
The Netherlands, 20-25 June 1999.
}
\end{center}

\thanks{
Work Supported by
Department of Energy contract  DE--AC03--76SF00515 (SLAC)
}

\begin{abstract}

During 1996, the SLD collaboration completed construction and began operation
of a new charge-coupled device (CCD) vertex detector (VXD3).
Since then, its performance has been studied in detail
and a new topological vertexing technique has been developed.
In this paper, we discuss the design of VXD3,
procedures for aligning it, and the tracking and vertexing improvements
that have led to its world-record performance.

\end{abstract}

\end{frontmatter}

\normalsize
 
 
%
\pagestyle{plain}

\section{Introduction}               

A technique of charge-coupled device (CCD) detectors
for high energy experiments
was established by the NA32 fixed-target experiment at CERN 
in mid 80s~\cite{Damerell:1986rd}.
They used a CCD-based vertex detector for the identification of
charmed particles.
Their results showed excellent vertexing performance for
short-lived particles.
It was realized that these devices offered
the possibility of excellent physics performance in the $e^+e^-$
linear collider environment.
The SLD experiment at the SLAC Linear Collider (SLC) is the first
experiment which uses CCDs as a vertex detector in a colliding beam
experiment.
After tests with a prototype detector VXD1, the 120M pixel detector
VXD2~\cite{Damerell:1990ie,Damerell:1992fb}
was installed for physics runs starting in January 1992.
During the SLD runs with VXD2, we developed a topological vertex finding
algorithm~\cite{Jackson:1997sy} to tag heavy-quark jets.
The very small ($\mu\mathrm{m}$ size) and stable SLC interaction point (IP)
and cleanly and precisely measured space points ($\sim 5.4 \mu\mathrm{m}$)
in VXD2 permit efficient identification of secondary 
and even tertiary vertices.
With VXD2 and this technique, we enjoyed
an advantage in both $b$- and $c$-jet tagging~\cite{Abe:1997sb}
over other collider experiments.

Rapid advances in CCD technology over the past 10 years made it possible
to replace VXD2 with a much more powerful 
vertex detector.
The upgraded vertex detector is called VXD3~\cite{Abe:1997bu} 
and was installed in December 1995.
VXD3 provides much better impact parameter resolution, 
larger solid angle coverage, and virtually error-free track linking.
All these features enhance the SLD heavy-quark measurements.
One of the most exciting possibilities is the search for $B^0_s$ mixing,
leading to the measurement of the oscillation frequency, $\Delta m_s$, 
and an improved determination of the CKM matrix element, $V_{td}$.
The initial performance of VXD3 can be found in 
Refs.~\cite{Abe:1997bu,Sinev:1997rj,Brau:1998ec}.
Since then, significant improvements in alignment and tracking algorithms
have led to remarkable performance results.

In this paper, we report the current performance of the SLD VXD3.
The features of CCD vertex detectors and VXD3 are described 
in Section~\ref{Sec:VXD3}.
Section~\ref{Sec:Alignment} discusses the alignment correction 
for precise space point determination and current achieved
spatial resolution.
In Section~\ref{Sec:SolidAngleCoverage}, we present
the VXD3-aided track finder which improves the solid angle coverage
of track reconstruction at SLD.
Section~\ref{Sec:GhostTrackAlgorithm} describes the track-fitter improvement
and a new topological vertexing technique.
Our latest impact parameter resolution and topological vertexing performances
are also shown in this section.
Finally a summary is given in Section~\ref{Sec:Summary}.

\section{The SLD VXD3 \label{Sec:VXD3} } 

CCD-based vertex detectors are well matched to $e^+e^-$ linear colliders,
providing nearly ideal experimental conditions for heavy flavor physics,
for the following reasons:
\begin{enumerate}
\item	Very small beam spots ($\mu\mathrm{m}$ size), 
hence a well defined primary vertex for every event.
\item	Highly segmented pixel structure, which provides natural 
3-dimensional space points and would comfortably absorb
high background per bunch crossing, likely to be found in a linear collider.
\item	Precise space-point resolution, 
resulting to date in a measurement precision of 
$< 4\mu\mathrm{m}$ in space points.
\item	Very thin detectors and small beam pipe radius, hence
degradation of impact parameter resolution
due to multiple scattering could be greatly reduced.
\item	Long interval between bunch crossings.
While this is not sufficient for complete readout, 
the average integrated background 
during readout was only $\sim 10$ bunch crossings.
\end{enumerate}

\begin{figure}
\centerline{
	\epsfxsize 5.4 truein 
	\epsfbox{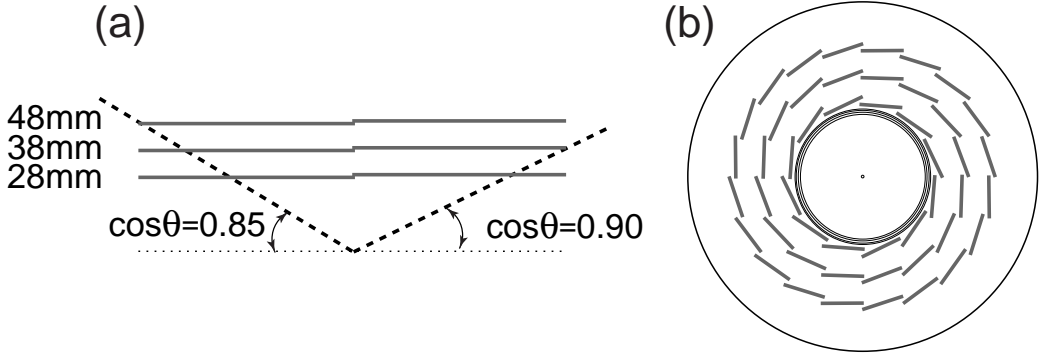}
}
\caption{
Layout of the CCDs in VXD3, 
(a) in views along the beam direction and (b) transverse.
}
\label{Fig:layout-of-ccds-in-vxd3}
\end{figure}

The SLD experiment is the only collider experiment which satisfies the above
conditions.
The detailed description of VXD3 is found in Ref.~\cite{Abe:1997bu}.
Using advances in CCD technology, in particular increasing the device active
area, VXD3 has the following features:
\begin{enumerate}
\item	Extended polar angle coverage, to benefit from the large polarized
asymmetry in physics processes in the most valuable regions of high 
$|\cos\theta|$.
\item	Full azimuthal coverage in each of three barrels, to achieve
redundancy and self-tracking capability independent of the Central Drift
Chamber (CDC), 
and consequently improved overall tracking efficiency.
\item	Optimized geometry with stretched radial lever arm and reduced
material in each layer, for significantly improved impact parameter resolution.
\end{enumerate}
VXD3 consists of 96 CCDs~\cite{Ref:EEV} arranged on 3 cylindrical layers 
of beryllium supporting ladders around the interaction point (IP).
Only 2 CCDs cover the entire length of the 159mm ladder 
with an overlap of about 1mm in the region near $z=0$, 
see Fig.~\ref{Fig:layout-of-ccds-in-vxd3}a.
Ladders in the same layer are placed in a `shingled' layout with
a small cant angle of $9-10^{\circ}$.
The layout provides azimuthal coverage overlap in the range of 
$300\mu\mathrm{m}$ to 1mm, depending on layer and CCD location,
see Fig.~\ref{Fig:layout-of-ccds-in-vxd3}b.
The overlaps allow an internal detector alignment by using the tracks
passing through the overlap regions,
discussed in Section~\ref{Sec:Alignment}.
With a beam pipe inner radius of 23.5mm and the layer-1 radius of 28.0mm,
the layer-3 radius of 48.3mm achieves complete azimuthal coverage out to
$|\cos\theta|<0.85$ for $\geq 3$ VXD hits
and provides  enough lever arm for an accurate measurement of track angles.
If 2 VXD hits are permitted, the layer-2 radius of 38.2mm allows
an extend coverage of $|\cos\theta|<0.90$ 
(see Section~\ref{Sec:SolidAngleCoverage}).

The CCDs are n-buried channel devices fabricated on p-type epitaxial layer 
and having a p+ substrate.  
The transverse pixel sizes are $20\mu\mathrm{m} \times 20\mu\mathrm{m}$, 
and the epitaxial layer is 20 $\mu\mathrm{m}$ thick.
They have an active area of $80\mathrm{mm} \times 16\mathrm{mm}$. 
Each CCD contains $4000 \times 800 = 3.2\mathrm{M}$ pixels,
so there are $3.2\mathrm{M} \times 96=307\mathrm{M}$ pixels
in VXD3.
They are operated in a full-frame readout mode.
The readout register operates on 
two-phase clocking and the imaging area on three-phase.  
There are 4 readout nodes, one on each corner of each CCD device, 
and the pixel readout rate is 5 MHz. 
With beryllium motherboard stiffeners and the CCDs thinned to 
$150\mu\mathrm{m}$, 
a VXD3 ladder is $0.4\mathrm{\%}$ radiation lengths thick. 
Extended layer to layer distance arms and reduced detector material improves 
impact parameter resolution significantly compared to VXD2, 
as discussed in Section~\ref{Sec:GhostTrackAlgorithm}.
%
\section{Alignment and spatial resolution \label{Sec:Alignment} } 
%
We expected VXD3 to give an excellent intrinsic hit resolution 
of $\sim 4\mu\mathrm{m}$.
However, without a detailed internal alignment of the detector as installed 
in SLD, the average position resolution was about 4 times worse 
($> 16\mu\mathrm{m}$),
based on an initial optical survey of VXD3 performed during its
assembly.
This indicates that a precise calibration is needed to achieve
the full physics potential of the device.
The effective single hit resolution is the product of the 
intrinsic resolution and the systematic uncertainties 
in the spatial locations and orientations of the CCDs themselves.
The aim of the internal alignment is to remove the latter contribution.
In order to remove the uncertainty due to the
orientations of the CCDs, we use the charged track data itself.
This section describes the techniques used to
align the CCDs and obtain design performance~\cite{Ref:Alignment}.

In order to correct the alignment systematics,
we introduce 6 parameters per CCD as a first order correction.
The parameters consist of three translations and three rotations.
There are a total of $6 \times 96$ corrections
for these degrees of freedom.
In addition, a pair of parameters are introduced for the deviation of 
the interaction point from the nominal. 

During the assembly of VXD3,
the relative positions of the CCDs within each barrel of VXD3 were measured
with typically a few 10s of $\mu\mathrm{m}$ 
precision in a room-temperature optical survey.
This survey geometry forms the starting point
with which the first data was reconstructed.   
The alignment procedure assumed that each CCD was approximately flat, 
with small shape corrections as measured in the optical survey.

Misalignments of CCDs cause a measured hit on a CCD
to be displaced from a true track trajectory by a residual $\delta$.
Even though we correct the translations, the rotations and 
the initial CCD shape,
higher order systematics exist in the track residual distributions.
We ascribe these systematics to CCD shape changes
after the installation, cooling and/or imperfections 
in the original optical survey.
These deviations can be approximated by 
a 4th-order polynomial along the $z$ direction.
Fixing the deviations at the CCD edges to be 0,
we introduce 3 additional alignment parameters per CCD to correct 
these higher order systematics.
The alignment parameters significantly correct the radial position of
a hit point on the CCD which is particularly important 
at large $|\cos\theta|$.
Now we have 9 alignment parameters per CCD and there are
$9 \times 96 + 2 = 866$ alignment parameters in total.

Since the true track trajectory is unknown
it is necessary to identify hits on at least three CCDs associated
with a track reconstructed in CDC.
We use only the measured momentum from the CDC track to constrain 
the track extrapolation inside VXD3.
Good quality tracks are selected with a momentum of at least 1~GeV.
In general the track is constrained to pass through two of the CCD
hits and the corresponding residual measured the third, reference,
CCD.
According to combinations of three of CCDs, we classify groups of residuals,
shingles, doublets, triplets, pairs and so on~\cite{Ref:residualtype}.
The small curvature effect of the charged track
in the SLD magnetic field is taken into account in the
$r\phi$ plane.

For each type of residual and each unique combination of CCDs 
data files are
accumulated containing the deviations $\delta$ in $r\phi$ and $rz$
as a function of $\tan\lambda$ and $\phi$, 
where $\tan\lambda$ is a dip angle.
These data files are fitted to the function forms such as 
Eq.~\ref{Eq:residual_fit_parameters},
to determine the parameters of the deviations ($C_1$, $C_2$, $\ldots$) 
and the error matrix of these parameters.
\begin{equation}
  \delta = C_1 + C_2  \tan\lambda + C_3 \tan^2\lambda \ldots
\label{Eq:residual_fit_parameters}
\end{equation}
These fits are done using MINUIT~\cite{James:1975dr} 
with an automated procedure 
to loop over the large number (${\sim}1500$) of
residual type/CCD combinations involved. 

\begin{figure}
\centerline{
	\epsfysize 3.3 truein 
	\epsfbox{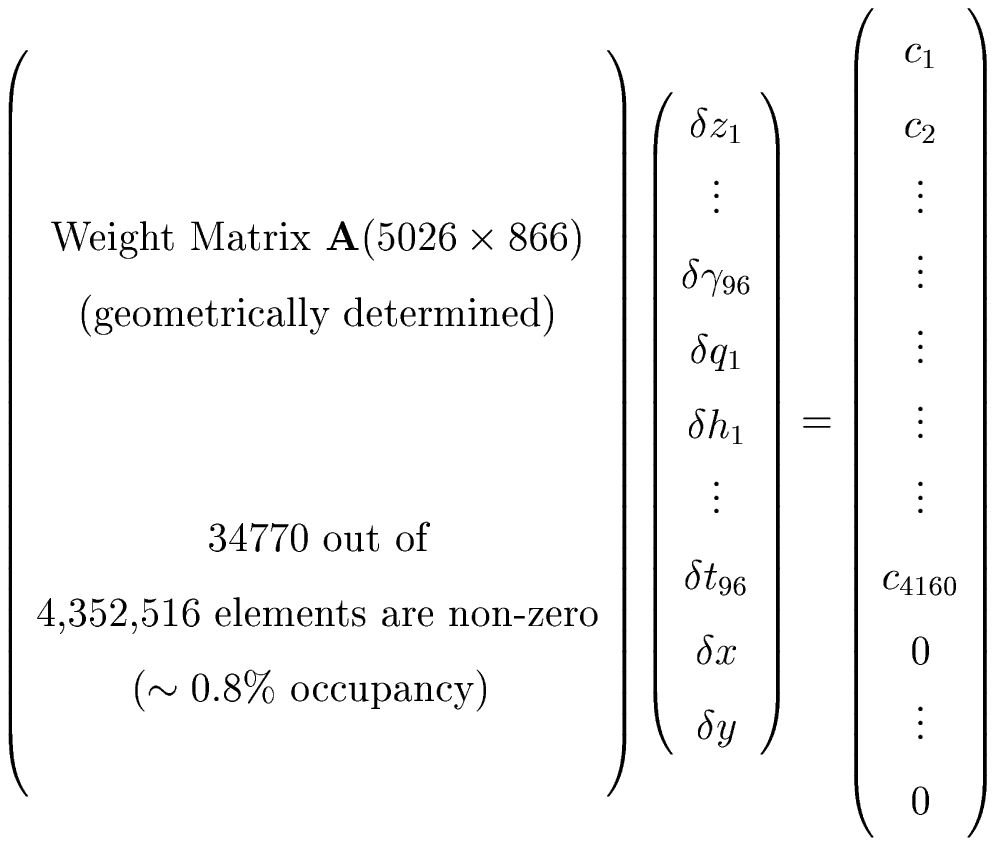}
}
\caption{
Schematic of the alignment matrix equation.
}
\label{Fig:AlignmentMatrix}
\end{figure}

By inspection it is now possible for each residual type
  to build matrices of the 
form \textbf{A  x = c} where \textbf{x} is a column matrix of the degrees of
freedom, \textbf{c} is a column matrix of parameters from the functional forms
of the deviations and \textbf{A} is a weight matrix determined by the nominal
geometry. 
Clearly the vector \textbf{x} must be common to all classes, 
since this is simply the corrections to the CCD positions needed 
to optimize the geometry, 
so that the matrices from the different classes can be combined 
as illustrated in Fig.~\ref{Fig:AlignmentMatrix}.
The values of alignment parameters can simply be obtained
by inverting the matrix \textbf{A}~\cite{Ref:SVD},
\begin{equation}
  \mathbf{x} = \mathbf{A}^{-1}\mathbf{c}.
\end{equation}

\begin{figure}
\centerline{
	\epsfxsize 5.4 truein 
	\epsfbox{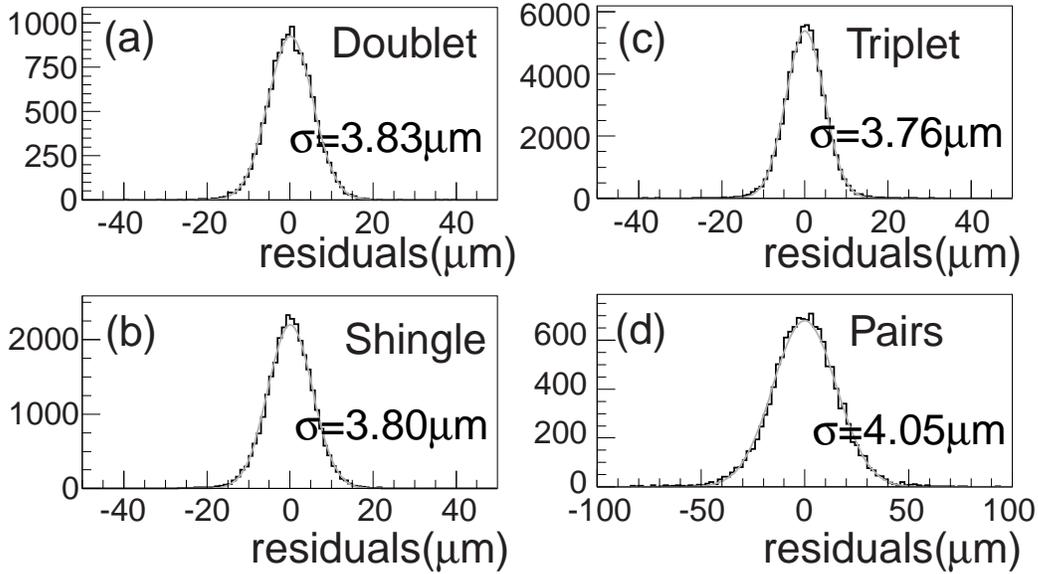}
}
\caption{
Residual distributions for
(a) doublet, (b) shingle, (c) triplet and (d) pairs.
The definition of these can be found in Ref.~\cite{Ref:residualtype}.
Fitting a Gaussian curve to each of these figures yields a 
width of $5.42\mu\mathrm{m}$ ,$5.37\mu\mathrm{m}$, $4.61\mu\mathrm{m}$
and $15.27\mu\mathrm{m}$ for doublets, shingles, triplets and pairs,
respectively, corresponding to spatial resolutions of 
$3.83\mu\mathrm{m}$, $3.80\mu\mathrm{m}$, $3.76\mu\mathrm{m}$ and
$4.05\mu\mathrm{m}$.
One hit resolution is consistently $\sim 3.8\mu\mathrm{m}$.
}
\label{Fig:residual_after_alignment}
\end{figure}

Fig.~\ref{Fig:residual_after_alignment} 
shows an example of the detector resolution
obtained with the final geometry using all events recorded 
in the $1997-98$ data.
From these results, we obtain a consistent one-hit resolution of 
$\sim 3.8\mu\mathrm{m}$.

It should be noted that this procedure does not need to 
be iterated. 
This is expected given the relatively minor nature of the
approximations made in the analysis and is confirmed to be correct for the
VXD3 geometry during 1996 data taking where a second iteration of the 
procedure produce no significant change to the geometry.

\section{Improved solid angle coverage \label{Sec:SolidAngleCoverage} } 

Many of the SLD physics analyses (particularly the $b$ mixing and fermion
asymmetry measurements) can benefit from an extended polar angle coverage,
due to the large analyzing power at high $|\cos\theta|$.

The tracking strategy adopted for VXD2 was to reconstruct
the CDC track first,
extrapolate it to the VXD and then search for the best VXD-hit combination
to form the complete track.
However, Monte Carlo studies indicate that around 5\% of the prompt tracks in
a hadronic $Z^0$ decay within the CDC active volume are either not tracked by
CDC or fail linking to the VXD, mainly due to contamination of wrong CDC hits
distorting the extrapolation.
This is a consequence of track merging in a dense hadronic jet environment,
compounded by the relatively small CDC outer radius and a rather moderate 
solenoid field of 0.6 Tesla.
This problem becomes progressively worse once the track $|\cos\theta|$ 
increases beyond 0.7, when the available tracking length is shortened.
Actually the linking efficiency with $\geq$2 VXD hits begins to decrease
at $|\cos\theta|=0.7$ and falls to $50\mathrm{\%}$ at $|\cos\theta|=0.75$.
VXD2 could not offer independent
assistance to relieve these pattern recognition deficiencies.

VXD3 was designed to have a full 3 layer coverage
as sketched in Fig~\ref{Fig:layout-of-ccds-in-vxd3},
allowing a self tracking within VXD3 alone.
For the majority of VXD3 tracks with $\geq 3$ hits,
a VXD-hit vector is relatively easy to reconstruct stand alone, 
while fake combinations only amount to less than $10\mathrm{\%}$
of all vectors before any matching with CDC tracks.
The VXD-hit vectors have the following nice features:
\begin{enumerate}
\item	High precision vectors:
The very fine granularity of the CCD pixels is ideal for resolving
the otherwise difficult merging track cases and discrimination of tracks
which are case together.
These high precision VXD-hit vectors in 3-D are powerful additions to
the global tracking pattern recognition capability.
\item	Wide acceptance:
A full azimuthal and extended polar angle coverage of VXD3 guarantees 
to  form VXD-hit vectors widely with high and flat
reconstruction efficiency. 
For $\geq$3-VXD-hit vectors, 
they can be reconstruct out to $\cos\theta=0.85$, beyond the VXD2 acceptance
of $|\cos\theta|<0.75$.
\end{enumerate}
The VXD-hit vectors are hence quite useful in improving 
the tracking performance in the forward region.

The new tracking strategy adopted for VXD3 uses 
vertex-detector-hit vectors at
the earliest stage of the track finding algorithm.
The global tracking pattern recognition combines both VXD- and CDC-hit 
vectors into a set of tracks.
Due to the 3 dimensional nature and fine segmentation of CCD pixel device,
the real combinations of VXD- and CDC-hit vectors stand out clearly
and fakes are easily rejected with a $\chi^2$ test.
To form complete tracks, a joint fit the selected vectors is performed 
as described in Section~\ref{Sec:GhostTrackAlgorithm}.
Some track are lost in the above procedure because of a kink
at the boundary between VXD3 and CDC.
In order to recover these tracks,
CDC-alone tracks are also reconstructed and extrapolated to the VXD3
to search for the best remaining VXD-hit combination.

Since the new tracking algorithm works well with 3-VXD-hit vectors,
we extend it to a more aggressive prospect, 2-VXD-hit vectors.
Using the 2-VXD-hit vectors, 
we get an angular range of $0.85<|\cos\theta|<0.9$, 
beyond the coverage of 3-VXD-hit vectors.
However the purity of 2-VXD-hit vectors 
($= \mbox{number of signal vectors}/\mbox{number of total vectors}$)
is not so high, 
and depends on a beam condition.
In 1997 run, the purity of 2-VXD-hit vectors is typically $35\mathrm{\%}$, 
while that for 3-VXD-hit vectors is $93\mathrm{\%}$.
The low purity results in a rather low purity of reconstructed tracks
and increased CPU time.

\begin{figure}
\centerline{
	\epsfysize 3.5 truein 
	\epsfbox{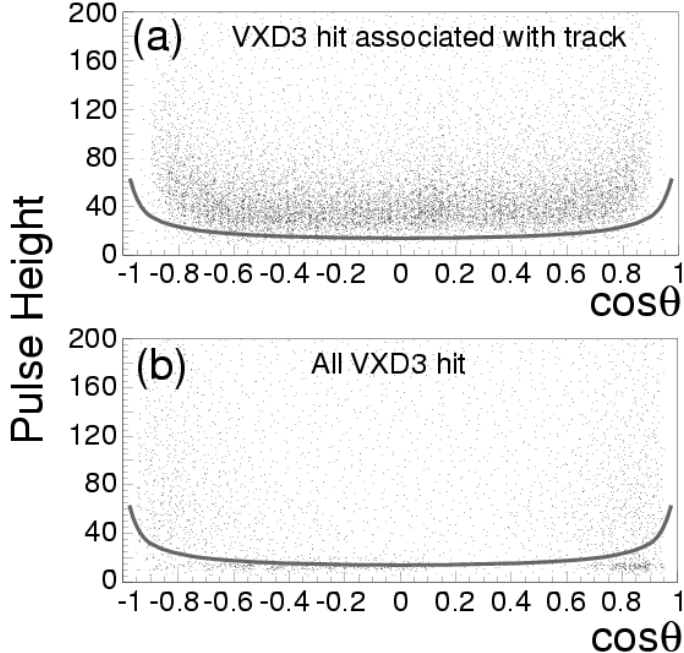}
}
\caption{
Pulse height distributions as a function of $\cos\theta$ for
(a) VXD3 clusters associated with tracks and
(b) all VXD3 clusters.
The lines indicate pulse height cut requiring $>14/\sin\theta$.
Signal pulse heights are above the line, while noise ones bunch 
below the line.
}
\label{Fig:PulseHeight}
\end{figure}

In order to improve the purity, we look at the pulse height information of 
clusters in 2-VXD-hit vectors.
We assume that the pulse heights depend on the path length of 
the charged particle in the CCD:
\begin{equation}
 \mbox{Pulse Height} \propto \frac{\mbox{Thickness of CCD}}{\sin\theta} .
\end{equation}
Fig.~\ref{Fig:PulseHeight} illustrates the pulse height distributions versus
$\cos\theta$ for VXD clusters associated with tracks and for all VXD clusters.
According to the results of Fig.~\ref{Fig:PulseHeight},
we require the pulse height of VXD cluster to be greater than $14/\sin\theta$.
The cut removes half of the fake 2-VXD-hit vectors.
Once the purity is improved, 2-VXD-hit vectors significantly improve
the forward tracking.
The detailed internal alignment and the high spatial resolution 
allow the 2-VXD-hit vectors to determine the track direction and $z$ position
accurately.
In the region of $0.85<|\cos\theta|<0.9$,
CDC has too few hits to determine these quantities reliably.
Therefore it is very difficult to reconstruct high quality tracks in
the forward region without 2-VXD-hit vectors.

\begin{figure}
\centerline{
	\epsfysize 3.5 truein 
	\epsfbox{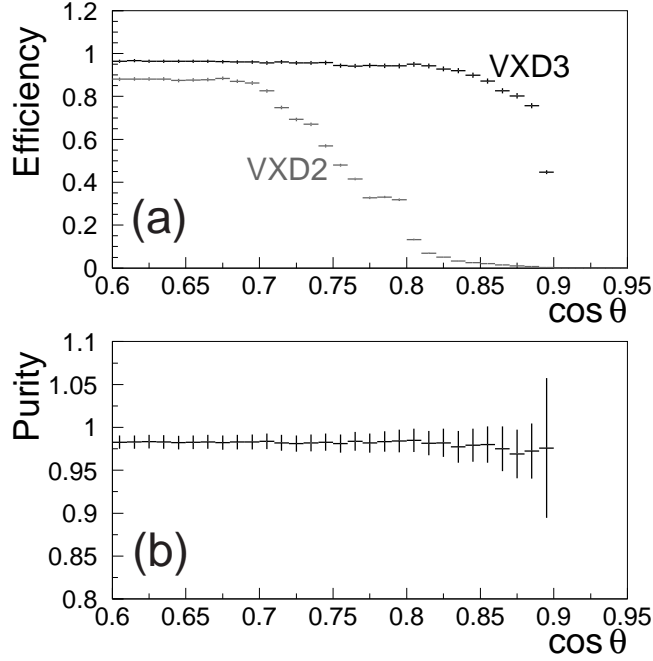}
}
\caption{
(a) The tracking efficiency curves as a function of $|\cos\theta|$
requiring $\geq$ 2 VXD hits.
Black point indicates new track reconstruction algorithm with VXD3 and
grey one original track reconstruction with VXD2.
(b) The track purity curve versus $|\cos\theta|$ with new track reconstruction
algorithm and VXD3.
}
\label{Fig:TUAC}
\end{figure}

Detailed Monte Carlo simulations are used to test the tracking performance.
Fig.~\ref{Fig:TUAC}a shows tracking efficiencies with $\geq2$ VXD hits
using the new track reconstruction algorithm with VXD3 and original 
reconstruction with VXD2, in hadron events.
Significant improvement can be seen in the region of $|\cos\theta|>0.7$
and the efficiency is flat to $|\cos\theta|=0.85$ and reasonably well
to near 0.9.
The track purity curve is shown in Fig.~\ref{Fig:TUAC}b.
The curve is uniformly high out to $|\cos\theta|=0.9$.
%

\section{Track fitter improvement and new topological vertexing technique 
		\label{Sec:GhostTrackAlgorithm} } 

\begin{figure}
\centerline{
	\epsfysize 3.5 truein 
	\epsfbox{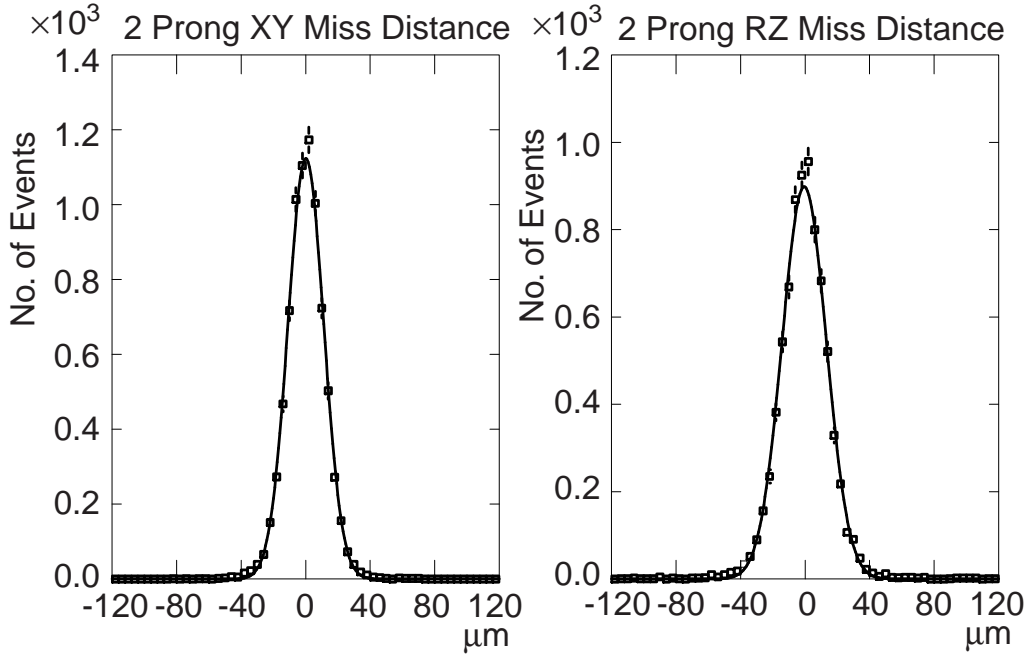}
}
\caption{
$\mu$-pair miss distances in $r\phi$ and $rz$ projections.
The fitted $\sigma$'s correspond to impact parameter resolutions of
$7.8\mu\mathrm{m}$ and $9.7\mu\mathrm{m}$.
Comparing the old impact parameter resolutions of
$10.7\mu\mathrm{m}$ ($23.5\mu\mathrm{m}$) in $r\phi$ ($rz$),
significant improvement is seen in $\sigma_{rz}$. 
}
\label{Fig:misdistance}
\end{figure}

Track fitting procedures have also been modified.
We introduce a new track extrapolation program 
to give more precise track reconstruction and a Kalman filter 
algorithm~\cite{Regler:1990an,Fruhwirth:1990py,Billoir:1984mz}
for CDC+VXD fit $\chi^2$ calculation
taking into account the effect of multiple scattering in detector material
more accurately than before.
During this process,
we also discovered, and corrected, that
the VXD3 cluster errors were not correctly assigned in the tracking code.
Fig.~\ref{Fig:misdistance} shows $\mu$-pair miss distance distributions 
in $r\phi$ and $rz$ projections.
Fitting a Gaussian curve to each figure yields a 
width of $11.0\mu\mathrm{m}$ and $13.7\mu\mathrm{m}$
for $r\phi$ and $rz$ projections respectively. 
These numbers are divided by the geometric factor $\sqrt{2}$ and
obtain the impact parameter resolutions of
\begin{equation}
\begin{array}{lll}
\sigma_{r\phi}=7.8\mu\mathrm{m} & \sigma_{rz}=9.7\mu\mathrm{m} &
	\mbox{ (new track fitter with VXD3)}\\
\sigma_{r\phi}=10.7\mu\mathrm{m} & \sigma_{rz}=23.5\mu\mathrm{m} & 
	\mbox{(original track fitter with VXD3)} \\
\sigma_{r\phi}=11\mu\mathrm{m} & \sigma_{rz}=38\mu\mathrm{m} &
	\mbox{(original track fitter with VXD2)} . \\
\end{array}
\nonumber
\end{equation}
Here impact parameter resolutions by original track fitter 
with VXD3 and with VXD2 are also shown.
With the detailed internal alignment, significant improvement can be seen, 
in particular $rz$ projections because
we put more weight on VXD3 than CDC which
has worse position resolution in $z$ direction, in the track fitter.
Current impact parameter resolutions as a function of momentum and
angle with VXD3 are:
\begin{equation}
\sigma_{r\phi} = 7.8 \oplus \frac{33}{p\sin^{3/2}\theta} \; \mu\mathrm{m}
\end{equation}
\begin{equation}
\sigma_{rz} = 9.7 \oplus \frac{33}{p\sin^{3/2}\theta} \; \mu\mathrm{m} .
\end{equation}
Our impact parameter resolution is outstanding compared with 
the other collider experiments.

Taking advantage of this significant improvement, we have developed
a new algorithm to reconstruct $B$ decay.
The new algorithm relies on the long $B$ and $D$ lifetimes and 
the kinematic fact that the large boost of 
the $B$ decay system carries the
cascade charm decay downstream from the $B$ decay vertex. 
Monte Carlo studies show that in $B$ decays producing a single $D$ meson
the cascade $D$ decays on average $4200\mu\mathrm{m}$
from the IP, while the intermediate $B$ vertex is displaced on average only
$46\mu\mathrm{m}$ transversely from the line joining the IP to the $D$ decay
vertex. 
This kinematic stretching of the $B$ decay chain into 
an approximately straight line is exploited 
by what is called the ghost track algorithm~\cite{Ref:GhostTrackAlgorithm}. 
This new algorithm has two stages and operates on a given set of 
selected tracks in a jet or hemisphere. 
Firstly, the best estimate of the straight line from the IP 
directed along the $B$ decay chain is found. 
This line is promoted to the status of a track by assigning it a 
finite width. 
This new track, regarded as
the resurrected image of the decayed $B$ hadron, is called the 
`\textsf{ghost}' track.
Secondly, the selected tracks are vertexed with the `\textsf{ghost}' track
and the IP to build up the decay chain along the ghost direction. 
Both stages are now described in more detail.

\begin{figure}
\centerline{
	\epsfysize 3.2 truein 
	\epsfbox{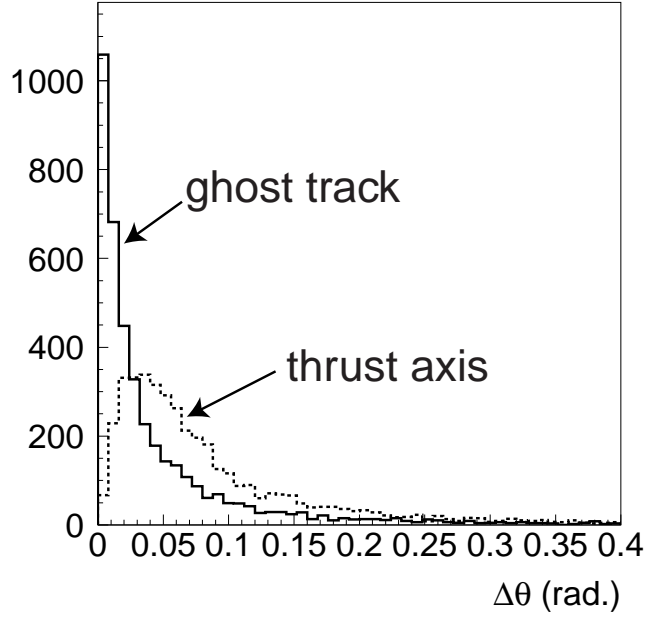}
}
\caption{
Residual distributions between a true $B$ direction and
a `\textsf{ghost}' track/thrust axis.
Open histogram indicates the distribution for a `\textsf{ghost}' track and
hashed one for a thrust axis.
The `\textsf{ghost}' track gives a better $B$ direction estimate
than the thrust axis.
}
\label{Fig:b-direction-resolution}
\end{figure}

A new track \textsf{G} is created 
with a set of tracks in a given hadronic jet or hemisphere.
Initially the track \textsf{G} is identical to the jet or thrust axis
and has a constant resolution width in both $r\phi$ and $rz$.
For each track $i$, a vertex is formed with 
the track \textsf{G} and the vertex location $\vec{r}_i$, fit
$\chi^2_i$ and $L_i$ are determined
($L_i$ is the longitudinal displacement from the IP to $\vec{r}_i$ 
projected onto the direction of track \textsf{G}). 
This is calculated for
each of the tracks and the summed $\chi^2_S$ is formed
to construct, such that 
when the direction of \textsf{G} is varied the minimum of $\chi^2_S$ provides 
the best estimate of the $B$ decay direction. 
After finding the minimum of $\chi^2_S$,
the width of track \textsf{G} is set such that the maximum $\chi^2_i = 1.0$
for all potential $B$ decay candidate tracks (L$_i > 0$).
The obtained track \textsf{G} is now called the `\textsf{ghost}' track.
Fig.~\ref{Fig:b-direction-resolution} shows difference between a true $B$
direction and a `\textsf{ghost}' track/thrust axis.
The `\textsf{ghost}' track gives a better $B$ direction estimate 
than the thrust axis because of the excellent impact parameter resolution.

The second stage of the algorithm begins by defining a fit probability
for a set of tracks to form a vertex with each other and with the
`\textsf{ghost}' track (or IP). 
This probability then measures the likelihood 
of the set of tracks both belonging to a common vertex \textbf{and}
being consistent with the `\textsf{ghost}' track (or IP) and hence forming 
a part of the $B$ decay chain. 
These probabilities are determined from
the fit $\chi^2$ which is in turn determined algebraically from the
parameters of the selected tracks and the `\textsf{ghost}' track 
(or the IP). 
Fake vertices peak at probability close to 0. 

The aim is now to find the most probable track-vertex associations
to divide the set of tracks and IP into subsets of reconstructed vertices.
Candidate vertices are groups of 1 or more tracks (or IP)
considered as the reconstruction develops.
For a set of N tracks, there are initially N+1 candidate
vertices (N 1-prong secondary vertices and a bare IP).
Fit probabilities for all pairs of candidate vertices are calculated
together with the `\textsf{ghost}' track.
The probabilities of each track associated with the IP are calculated too.
The pair of candidate vertices which have the highest probability
for fitting in a vertex together is found, 
and then combined to form a new candidate vertex.
This modifies the set of all candidate vertices, 
and the procedure then repeats with the new set.
At each iteration, 
the number of candidate vertices decreases by one.
The iterations continue until the maximum probability is less than 1\%.
At this point the tracks and IP have been divided into
unique subsets by the associations thereby defining topological
vertices.

\begin{figure}
\centerline{
	\epsfysize 3.2 truein 
	\epsfbox{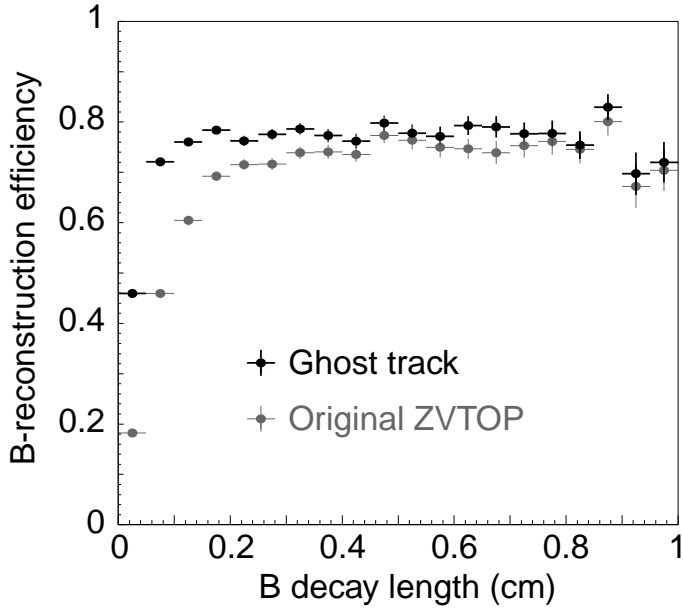}
}
\caption{
$B$ reconstruction
efficiency curves as a function of the $B$ decay length.
Black points indicate ghost track algorithm 
and grey points original topological vertexing algorithm.
The ghost track algorithm improves the efficiency along with entire $B$ decay
length, in particular short decay length.
}
\label{Fig:ghost-b-reconstrution-effieicney}
\end{figure}

The performance of the ghost track algorithm is quite remarkable.
Fig.~\ref{Fig:ghost-b-reconstrution-effieicney} shows $B$ reconstruction
efficiency curves as a function of the $B$ decay length.
The ghost track algorithm improves the efficiency along with entire $B$ decay
length, particularly at short decay length.
The $B$ reconstruction at the short decay length is very important 
because many $B$ hadrons decay in that region (statistical advantage) and
because proper time resolution (essential for $B_s$-mixing analysis)
is best at short decay lengths.

\begin{figure}
\centerline{
	\epsfysize 5.0 truein 
	\epsfbox{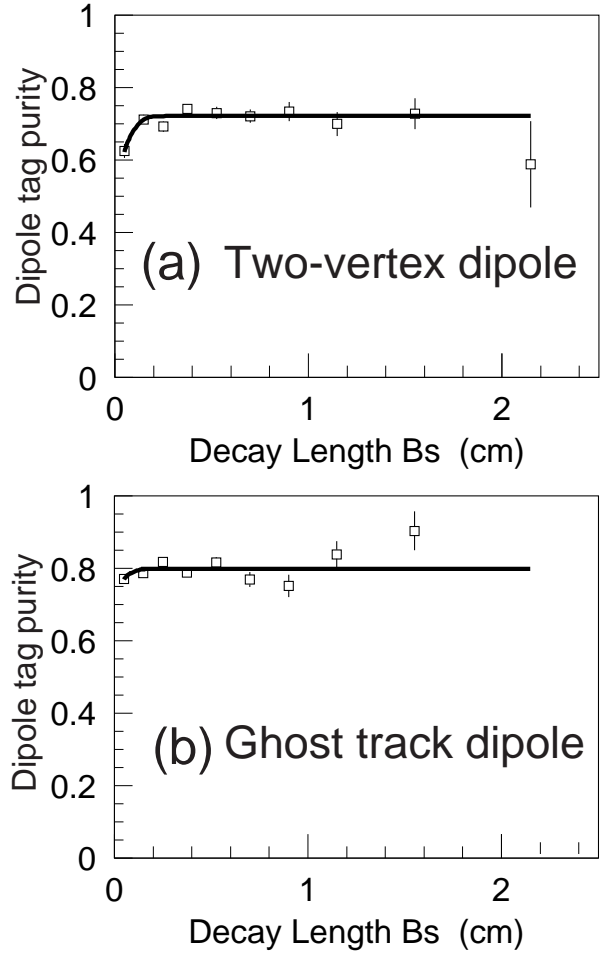}
}
\caption{
$B_s$-charge-dipole-tag purities dependent on the $B$ decay length for
(a) original topological vertexing algorithm and (b) the ghost track algorithm.
The ghost track algorithm increase purity for the entire region, 
in particular short decay length.
}
\label{Fig:dipole-tag-purity}
\end{figure}

The ghost track algorithm improves the $B$ vertex reconstruction purity, too.
Here we show the performance of $B_s$-charge-dipole-tag purity as an example.
Jets or hemispheres in which three vertices are found --
the primary (which includes by definition the IP), 
a secondary and a tertiary -- are used for the
charge-dipole analysis.
The secondary vertex is identified as the $B$
decay vertex and the tertiary as the cascade charm decay, 
and the charge dipole ($\delta q$) is defined as follows:
\begin{equation}
\delta q=d_{BD} \cdot \mathrm{sign} \left( Q_D - Q_B \right) .
\end{equation}
where $d_{BD}$ is the distance between charm and $B$ decay vertices,
$Q_D$ is the charge of $D$ vertex and $Q_B$ the charge of $B$ vertex.
Fig.~\ref{Fig:dipole-tag-purity} shows the charge-dipole-tag purity versus
$B_s$ decay length.
The ghost track algorithm improves the purity in the entire region,
again particularly at the short decay length.
The better purity implies that $B-D$ separation is improved, 
i.e. track attachment for the vertex is improved.
It should be mentioned that 
improving the purity and efficiency of $B$ reconstruction
(by requiring the vertices be consistent with a single line of flight),
the ghost track algorithm has the additional advantage of allowing
the direct reconstruction of 1-prong vertices, including the
topology consisting of 1-prong $B$ decay and $D$ decays.

\begin{figure}
\centerline{
	\epsfysize 5.0 truein 
	\epsfbox{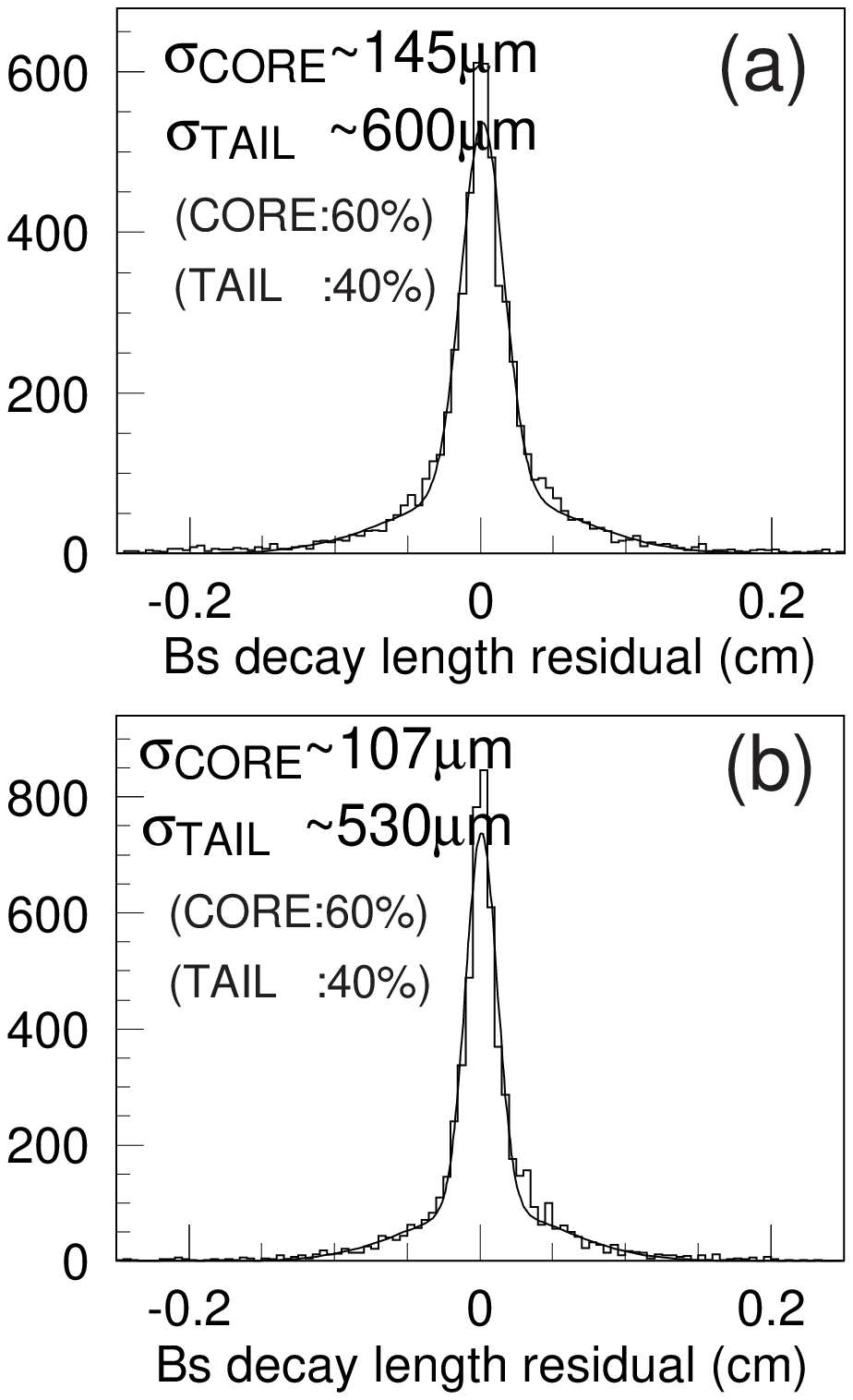}
}
\caption{
Residual distributions of $B_s$ decay length for charge-dipole-tag method.
(a) original track fitter plus original topological vertexing algorithm and
(b) new track fitter plus the ghost track algorithm.
The improved fitter and ghost track algorithm 
reduces the decay length resolution
by $\sim 50\mathrm{\%}$ and obtained resolution of $107\mu\mathrm{m}$.
}
\label{Fig:decay-length-resolution}
\end{figure}

As the last topic of this section,
we show $B$ decay length resolution as evidence of how well 
the improved track fitter plus the ghost track algorithm work.
Fig.~\ref{Fig:decay-length-resolution} illustrates that 
$B_s$-decay-length residual for an earlier dipole algorithm and 
the improved fitter with  the new ghost track algorithm.
The improved fitter and the ghost track algorithm  
reduce the core decay length resolution
by $\sim 50\mathrm{\%}$ obtaining a resolution of $107\mu\mathrm{m}$.
Even better resolutions are obtained in other analyses.
The best one is obtained by the reconstruction of 
$B_s \to D_s + X; D_s \to \phi+\pi$ mode.
In this case, the core resolution is $46\mu\mathrm{m}$.
This performance is also unsurpassed by the other experiments.
%

\section{Summary \label{Sec:Summary} } 

In this paper, we discussed
refinements in alignment, pattern recognition, and track fitting 
that have significantly improved combined CDC/VXD3 tracking performance.
The upgraded tracking performance has resulted in
world-record performance.
With improvements in topological vertexing technique,
it should lead to important physics results including greatly improved 
sensitivity for $B_s$ mixing.
The outstanding performance of VXD3 encourages us to aim toward even higher
performance goals for vertex detectors at future linear collider 
detectors~\cite{Burrows:1999ta}.



\begin{thebibliography}{9}            

\bibitem{Damerell:1986rd}
C.J.~Damerell, R.L.~English, A.R.~Gillman, A.L.~Lintern, F.J.~Wickens and S.J.~Watts,
IEEE Trans.\ Nucl.\ Sci.\ {\bf 33}, 51 (1986).

\bibitem{Damerell:1990ie}
C.J.~Damerell {\it et al.},
Nucl.\ Instrum.\ Meth.\ {\bf A288}, 236 (1990).

\bibitem{Damerell:1992fb}
C.J.~Damerell {\it et al.},
{\it  In *Dallas 1992, Proceedings, High energy physics, vol. 2* 1862-1866}.

\bibitem{Jackson:1997sy}
D.J.~Jackson,
Nucl. Instrum. Meth. {\bf A388}, 247 (1997).

\bibitem{Abe:1997sb}
K.~Abe {\it et al.}
[SLD Collaboration],
Phys.\ Rev.\ Lett.\ {\bf 80}, 660 (1998)
hep-ex/9708015.

\bibitem{Abe:1997bu}
K.~Abe {\it et al.},
Nucl. Instrum. Meth. {\bf A400}, 287 (1997).

\bibitem{Sinev:1997rj}
N.B.~Sinev {\it et al.}
[SLD Collaboration],
IEEE Trans. Nucl. Sci. {\bf 44}, 587 (1997).

\bibitem{Brau:1998ec}
J.E.~Brau
[SLD Collaboration],
Nucl. Instrum. Meth. {\bf A418}, 52 (1998).

\bibitem{Ref:EEV}
The CCDs were manufactured by the EEV Company, Chelmsford, Essex, UK.

\bibitem{Ref:Alignment} 
D.J.~Jackson {\em et al.}, 
to be submitted to Nucl. Inst. and Meth. {\bf A}.    

\bibitem{James:1975dr}
F.~James and M.~Roos,
Comput.\ Phys.\ Commun.\ {\bf 10}, 343 (1975).

\bibitem{Ref:residualtype}
The residual types in Fig.~\ref{Fig:residual_after_alignment} 
are defined as follows:
\begin{enumerate}
\item	Doublets: tracks through the overlapping region ($z$) of two CCDs
on the same ladder.
The vectors is fixed at one of the doublet hits and at the furthest-away
hit on a different layer.
The residual of the second doublet hit to this fixed vector is called 
the doublet residual.
\item	Shingles: tracks through the overlapping shingle region ($r\phi$)
of two ladders in the same layer.
The vector is fixed at the hit on one of the shingle ladders and 
at the furthest away hit on a different layer.
The residual of the hit from the second shingle ladder to the fixed vector
is called shingle residual.
\item	Triplets: tracks with hits in all three layers.
The vector is fixed at the Layer 1 and Layer 3 hits, and the Layer 2 hit
residual is called the triplet residual.
\item	Pairs: a pair of back-to-back particles. 
Each vector is given by hits in layers 1 and 3 (layer 2 is ignored for pairs).
The residual of the closest approach point between two vectors is called
pairs residual.
\end{enumerate}

\bibitem{Ref:SVD}
We use Singular Value Decomposition (SVD) technique 
in order to solve the equation.
A fuller discussion of the mathematical details can be found in
G. Golub and C. van Loan, Matrix Computations, (1983) John Hopkins University
Press.

\bibitem{Regler:1990an}
M.~Regler and R.~Fruhwirth,
\textit{In *St. Croix 1988, Proceedings, Techniques and concepts of high-energy
	physics* 407-499. }.

\bibitem{Fruhwirth:1990py}
R.~Fruhwirth, D.~Liko, W.~Mitaroff and M.~Regler,
\textit{ Presented at 5th Int. Conf. on Instrumentation for Colliding Beam
                  Physics, Novosibirsk, USSR, Mar 15-21, 1990}.

\bibitem{Billoir:1984mz}
P.~Billoir,
Nucl.\ Instr.\ Meth.\ {\bf A225}, 352 (1984).

\bibitem{Ref:GhostTrackAlgorithm}
D.J.~Jackson, 
\textit{reported at the International Europhysics Conference 
on High Energy Physics, Tampere, Finland, July 15-21, 1999}.
More detailed discussion can be found in SLAC-PUB-8225.

\bibitem{Burrows:1999ta}
P.N.~Burrows
[LCFI Collaboration],
these proceedings hep-ex/9908004.


\end{thebibliography}
\end{document}